\documentstyle[twocolumn,prl,aps,epsfig]{revtex}
\begin{document}                                                              
\draft
\title{Scattering of soft, cool pions}
\author{Robert D. Pisarski and Michel Tytgat}
\address{
Department of Physics, Brookhaven National Laboratory,
Upton, New York 11973-5000, USA}
\date{\today}
\maketitle
\begin{abstract}
Consider the effective lagrangian for pions in the chiral limit,
computed to leading order in an expansion about zero temperature.  
To describe the scattering of pions with small momenta,
it is necessary to include not just 
the usual shift in the pion decay constant, but also
a new, nonlocal term, which is precisely analogous to 
the hard thermal loops of hot gauge theories.
\end{abstract}
\pacs{BNL Preprint BNL-PT-961, Oct., 1996.}
\begin{narrowtext}

When a global symmetry is spontaneously broken, the interactions
between the Goldstone bosons follow uniquely by knowing
what the broken and unbroken symmetry groups are.
In hadronic physics
this gives chiral dynamics, which allows pion interactions 
to be computed by an expansion about low 
momenta\cite{chiral,wzw}.

For example, consider a theory like $QCD$, in which an exact 
(left-right) global
symmetry of $SU_\ell(N)\times SU_r(N)$ is
spontaneously broken to $SU(N)$.  
The chiral lagrangian is constructed from an unitary
$SU(N)$ matrix $g(x)$.
Writing down all terms invariant 
under global $SU_\ell(N) \times SU_r(N)$ rotations,
$g(x) \rightarrow U^\dagger_{\ell} g(x) U_{r}$,
there is only one term with two derivatives, 
\begin{equation}
{\cal L}_2 \; = \; {f_\pi^2\over 4}\; tr\; 
\left (\partial_\mu g^\dagger \partial_\mu g 
\right) \;\;\; , \;\;\; g^\dagger g = 1 \; .
\label{e1}
\end{equation}
Since the original symmetry is assumed to be exact, the 
$N^2-1$ Goldstone fields in $g$ are truly massless;
$f_\pi$ is the pion decay constant, 
$= 93$ MeV in $QCD$.  A systematic expansion is developed by
writing down all terms with increasing number
of derivatives \cite{chiral}; 
${\cal L}_2$ 
obviously dominates over terms with more than two derivatives
when the momenta in the $g$ 
field are small, less than $f_\pi$.

At first sight, it seems as if the extension of this analysis
to a nonzero temperature $T$ is elementary.
We limit ourselves to ``cool'' pions, where the 
temperature $T \ll f_\pi$,
and work to leading order in an expansion about zero temperature,
which is $\sim T^2/f_\pi^2$.  At such low temperatures it is
reasonable to compute the effects of a thermal bath using only
the chiral lagrangian.  (At higher temperatures $T \sim f_\pi$,
other modes become light, and eventually the full global symmetry
is restored.)  To this order the pion decay constant decreases 
as\cite{fpit}
\begin{equation}
f_\pi(T) = \left(1 - {N\over 24} {T^2\over f_\pi^2}\right) f_\pi \; .
\label{e2}
\end{equation}
This effect is incorporated in the chiral lagrangian
simply by replacing $f_\pi$ with $f_\pi(T)$.

Suppose, however, that we are interested in the scattering between
pions which are are not only cool but ``soft''.  
At nonzero temperature, 
in the imaginary time formalism (bosonic) scattering
processes are given by analytically continuing an amplitude,
computed as a function of the 
euclidean momentum $p^0 = 2 \pi m T$ for integral $m$, to
minkowski energies, $p^0 \rightarrow - i \omega + 0^+$.
We define soft to mean that each component of every pion
momenta is much less than the temperature: 
$|\omega|, |\vec{p}| \ll T \ll f_\pi$.

In this Letter we show that for the scattering of soft, cool
pions, the chiral lagrangian involves both ${\cal L}_2$
with $f_\pi(T)$ and a new, nonlocal term.
According to the standard power counting of momenta, this new term
is as important as the original ${\cal L}_2$.  While nonlocality
is unusual in an effective lagrangian, exactly the same type
of terms appear in perturbative analysis of 
gauge theories at high temperature,
where they are known as hard thermal loops\cite{htl}-$\!\!$\cite{eff}.

Hard thermal loops arise from scattering in a thermal bath between massless
fields, from processes corresponding to Landau damping.
The crucial part is that the fields have to be massless, so this
is natural either for gauge theories --- due to the gauge principle ---
or for Goldstone bosons --- because of Goldstone's theorem.
The rest is automatic, since scattering processes at nonzero
temperature always involve Landau damping.  In both
cases it is necessary to assume that the momenta are soft
so that hard thermal loops dominate over other, nonthermal, processes.

In principle our analysis applies to any system with
Goldstone bosons at nonzero temperature, such as for 
spin waves in ferromagnets and antiferromagnets.
We discuss later how hard thermal loops modify
scattering processes between Goldstone bosons.

In $QCD$, at low energies
$N=2$ (or $3$), but nonzero current quark masses imply
that the symmetry is approximate, and pions (or kaons and the
$\eta$) are massive.
Since $m_\pi \sim f_\pi$, the direct relevance of our results
for thermal pions can only be determined after detailed analysis;
they appear less applicable to kaons and the $\eta$.

To calculate an effective lagrangian we follow 
Polyakov and use the background field method
to take \cite{chiral,polyakov,ren} 
\begin{equation}
g \; \rightarrow \; g \; exp\left( i \pi/f_\pi \right) \; .
\label{e1a}
\end{equation}
After this redefinition, the new $g$ becomes a background field, which is
assumed to satisfy the equations of motion.  We expand
in fluctuations in $\pi$, 
$\pi = \pi^a \lambda^a$, with $\lambda^a$ the generators
of $SU(N)$, normalized as $tr(\lambda^a \lambda^b) = 2 \delta^{a b}$.
It is convenient to introduce the (right-handed) 
current $R_\mu(x)$,
\begin{equation}
R_\mu \; = \; g^\dagger \partial_\mu g \; .
\label{e9b}
\end{equation}
$R_\mu$ is like a gauge field in a $SU(N)$ gauge theory,
except that it is constrained to equal a pure gauge transformation
of the background field $g(x)$.
Consequently, the corresponding nonabelian field strength vanishes,
\begin{equation}
\partial_\mu R_\nu - \partial_\nu R_\mu + [R_\mu, R_\nu] = 0 \; .
\label{e9a}
\end{equation}
Note that the value of the gauge coupling constant $=1$.

Expanding the original action in powers of $\pi$, to linear
order we get the equation of motion for $g$, $\partial_\mu R_\mu = 0$,
which looks like the Landau gauge condition for $R_\mu$.  
To quadratic order,
\begin{equation}
{\cal L}_2 \rightarrow 
- \frac{f^2_\pi}{4} \; tr (R_\mu^2) 
+ {\cal L}_2(\pi) \; ,
\label{e3}
\end{equation}
\begin{equation}
{\cal L}_2(\pi) = \frac{1}{4} \; tr \left( (D_\mu \pi)^2 \right) 
- \frac{1}{16} \; tr \left( [R_\mu, \pi]^2 \right) \; ,
\label{ea4}
\end{equation}
where $D_\mu$ is the 
covariant derivative in the adjoint represention,
\begin{equation}
D_\mu = \partial_\mu + \frac{1}{2} [R_\mu,\cdot] \; ,
\label{e4}
\end{equation}
and $[\cdot , \cdot]$ denotes the commutator.
The first term in ${\cal L}_2(\pi)$ is the minimal coupling of 
a matter field $\pi$ to a gauge field $R_\mu$,
while the second term violates the gauge symmetry\cite{chiral,ren}.
The value of the
$SU(N)$ coupling constant in the covariant derivative
of (\ref{e4}) is not $=1$, 
as in (\ref{e9a}), though, but $=\frac{1}{2}$.
For coupling constant $=\frac{1}{2}$, $R_\mu$ is not 
pure gauge, and has nonzero field strength,
\begin{equation}
F_{\mu \nu} = 
\partial_\mu R_\nu \!- \!\partial_\nu R_\mu
 \!+ \!\frac{1}{2} [R_\mu, R_\nu]= 
\frac{1}{2} (\partial_\mu R_\nu \!- \!\partial_\nu R_\mu).
\label{e4b}
\end{equation}
Due to the detailed properties of the diagrams which
contribute to hard thermal loops, 
viewing $R_\mu$ as a gauge field will turn out to be a 
most fruitful analogy.  

We compute the effective lagrangian for $R_\mu$,
and so that for $g$, which
is generated by integrating out the $\pi$'s
to one loop order with ${\cal L}_2(\pi)$.  To do so, we follow the
analysis of gauge theories to isolate the dominant
terms between $R_\mu$ fields with soft momenta;
these are the hard thermal loops, so named because
the fields in the loop, here the $\pi$'s, have hard momenta, on the
order of the temperature\cite{htl}.

We start with the ``self energy'' $\Pi^{\mu \nu}(P)$ of $R_\mu$,
\begin{equation}
\Pi^{\mu \nu}(P) 
= \!\frac{N}{2} 
T \sum_{n}\!\!\int \!\!\frac{d^3 k}{(2 \pi)^3} 
 \frac{(2K + P)^\mu (2 K + P)^\nu}{K^2 (K + P)^2} \; .
\label{e5}
\end{equation}
where $P^\mu$ is the momentum of
the background field $R_\mu$, and the loop momentum
$K^\mu = (2 \pi n T, \vec{k})$, $n=-\infty$ to $+\infty$.
In a true gauge theory,
the gauge self energy is the sum of two diagrams:
one is similar to fig. [1], with two cubic vertices,
while the second diagram is a 
tadpole term from the quartic coupling.
\begin{figure}[hbt]
\centerline{\epsfig{figure=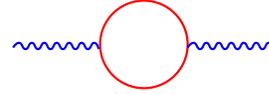,height=35mm,angle=90}}
\caption{Self energy of the $R_\mu$ field.}
\end{figure}

There is no quartic coupling in ${\cal L}_2(\pi)$, so
only the diagram of fig. [1] contributes; the
curly external lines denote $R_\mu$'s, the internal
solid lines $\pi$'s.
The overall factor of $N$ in $\Pi^{\mu \nu}(P)$ arises because 
each vertex involves a $SU(N)$ commutator,
and so, in component notation,
is proportional to the structure
constant $f^{a b c}$; thus fig. [1] is proportional to 
$f^{a c d} f^{b c d} = N \delta^{a b}$.
For arbitrary $P$ the complete integral in $\Pi^{\mu \nu}(P)$ is 
involved, but when $K$ is hard and $P$ soft,
in the numerator of the integral
we can neglect the terms $\sim K^\mu P^\nu$, $P^\mu K^\nu$,
and $P^\mu P^\nu$ relative to that $\sim K^\mu K^\nu$.
For this remaining integral we then isolate the term $\sim T^2$,
which defines the hard thermal loop \cite{htl}:
\begin{equation}
T\sum_{n}\!\!\int\!\!\frac{d^3 k}{(2 \pi)^3}  
\frac{K^\mu K^\nu}{K^2 (K+P)^2}\approx
\frac{T^2}{24}(\delta^{\mu \nu}\!-\!\delta\Pi^{\mu\nu}(P)).
\label{e6}
\end{equation}
In an effective lagrangian, the term $\sim \delta^{\mu \nu}$ 
changes the coefficient of $tr(R_\mu^2)$ from 
$f_\pi^2$ to $f^2_\pi(T)$ in (\ref{e2}).
There are several equivalent ways of writing
the effective lagrangian for 
hard thermal loops \cite{htl}-$\!\!$\cite{eff};
instead of following the original form of Taylor and
Wong\cite{taylor}, we include $\delta\Pi^{\mu \nu}$, and
go from momentum to coordinate space, by using\cite{eff}
\begin{equation}
T \sum_{m} \int \frac{d^3 p}{(2 \pi)^3} 
 \; tr \left(
R_\mu \; \delta \Pi^{\mu \nu}(P) \; R_\nu 
\right) = 
\label{e7}
\end{equation}
$$
\int \!\! d^4 x \!\! \int \!\! \frac{d \Omega_{ \hat{k} } }{4 \pi} 
\, tr \!\! \left(\!\! 
(\partial_\mu R_\alpha \!\! - \!\! \partial_\alpha R_\mu )
\frac{\hat{K}^\alpha\hat{K}^\beta}{- (\partial \cdot \hat{K})^2}
(\partial_\mu R_\beta \!\! - \!\! \partial_\beta R_\mu) \!\! \right) ,
$$
$\partial \cdot \hat{K} = \partial_\mu \hat{K}^\mu$.
We introduce a four vector $\hat{K}^\mu = (i,\hat{k})$, with
$\hat{k}$ a three vector of unit norm, $\hat{k}^2 = 1$, so
that $\hat{K}^\mu$ is null, $\hat{K}^2 = 0$.  This is a dummy
variable, in that one integrates over all directions of
$\hat{k}$, as $\int d \Omega_{\hat{k}}/(4 \pi)$.

In (\ref{e7}) $R_\mu$ appears as
$\partial_\mu R_\alpha-\partial_\alpha R_\mu$, {\it etc.},
because in momentum space $\delta\Pi^{\mu \nu}$
is transverse, $P^\mu \delta\Pi^{\mu \nu}(P) = 0$.
There is another way of seeing why this combination arises.
Assume that any hard thermal loop 
of the $R_\mu$'s is of the form
$\int d\Omega_{\hat{k}} (\partial_\alpha R_\beta)
\hat{K}^\delta\hat{K}^\gamma/(\partial \cdot\hat{K})^2 
(\partial_\lambda R_\kappa)$, and consider contracting the six
indices in all possible ways.  Using the equation of 
motion $\partial_\mu R_\mu = 0$, and the identity
$\int d\Omega_{\hat{k}}/(\partial \cdot \hat{K})^2 
=4 \pi/\partial^2$ \cite{eff}, 
one can show that all possible contractions reduce 
either to $tr(R_\mu^2)$ or to (\ref{e7}).

In (\ref{e6}) we compute
the hard thermal loop in the self energy of $R_\mu$,
but there are also hard thermal loops between three or
any higher number of $R_\mu$'s.  In a gauge theory,
because the cubic vertex can bring in a factor of
the hard loop momentum, while the
quartic vertex is independent of momentum,
the diagrams which contribute to
hard thermal loops with three or more gauge fields 
include only cubic, and never quartic, vertices \cite{htl}.
To sum up all such diagrams, as in fig. [2], we can then
neglect the second term in ${\cal L}_2(\pi)$,
and consider $\pi$ as a matter field in interaction with
a background gauge field $R_\mu$.  Once we are dealing with a gauge theory, 
however, the general form of the effective lagrangian is
uniquely determined by invoking gauge invariance,
in this instance for an adjoint field with coupling constant
$=\frac{1}{2}$.  To make (\ref{e7}) gauge invariant we merely replace
$\partial \cdot \hat{K}$ by $D \cdot \hat{K}$, and,
because of (\ref{e4b}),
$\partial_\mu R_\nu - \partial_\nu R_\mu$ by $2 F_{\mu \nu}$.
Hence to leading order in an expansion about zero temperature and momentum, 
$\sim (T^2/f_\pi^2) P^2$,
the dominant corrections to the chiral lagrangian are
given by $\delta {\cal L}_2$ \cite{foot}, 
\begin{equation}
{\cal L}_2 + \delta{\cal L}_2 =
- \frac{f^2_\pi(T)}{4} \; tr (R_\mu^2)
\label{e9}
\end{equation}
$$
- \frac{N T^2}{12} \int \frac{d \Omega_{ \hat{k} } }{4 \pi} \; 
tr \left(
F_{\mu \alpha}
\frac{ \hat{K}^\alpha \hat{K}^\beta}{- (D \cdot \hat{K})^2}
F_{\mu \beta}
\right) 
\; .
$$
If we naively power count momenta, then
$R_\mu \sim P_\mu$ and
$F_{\mu \nu} \sim P_\mu P_\nu$.
This shows that the reason why we can get a new term of
the same order as ${\cal L}_2 \sim P^2$ is because 
$\delta{\cal L}_2$ is nonlocal,
$\int F (1/\partial^2) F \sim
P^2 (1/P^2) P^2 \sim P^2$.

\begin{figure}[hbt]
\centerline{\epsfig{figure=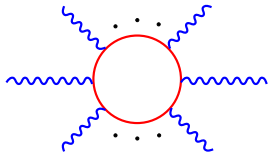,height=35mm,angle=90}}
\caption{Diagrams which contribute to $\delta{\cal L}_2$.}
\end{figure}

The nonlocal term in $\delta{\cal L}_2$ only matters for 
scattering processes where $\partial_0 g \neq 0$.
In the static limit, $\partial_0 g = 0$,
the nonlocal term in $\delta{\cal L}_2 \sim R_0^2$ \cite{eff}.
For static chiral fields $R_0 = 0$, and so
${\cal L}_2 + \delta{\cal L}_2$ 
reduces to $- f_\pi^2(T) tr(R_i^2)/4$.
Expanding $\delta{\cal L}_2$
in terms of pion fields, $g = exp(i \pi/f_\pi)$,
to lowest order 
$R_\mu \sim i \partial_\mu \pi/f_\pi$ and
$F_{\mu \nu} \sim [\partial_\mu \pi,\partial_\nu \pi]/f_\pi^2$.
Thus the nonlocal term in
$\delta{\cal L}_2$ first appears in scattering processes involving four
or more pions.

While the formal analogy between chiral and gauge fields
is very useful, the physics which follows is rather different.
The original lagrangian of the chiral field, $\sim tr( R_\mu^2 )$,
is for the gauge field a gauge variant mass term, 
and so doesn't arise.
Gauge fields have a hard thermal loop in their self energy,
which generates the screening of
electric and (time dependent) magnetic fields.  
For chiral fields, the nonlocal term
in $\delta{\cal L}_2$ doesn't affect
the propagation of pions, only their scattering.
Lastly, in gauge theories hard thermal loops are, for
soft momenta, as large as the terms at tree level;
consequently, a consistent perturbative expansion requires
their resummation through an effective expansion.
For soft, cool pions, on the other hand,
the hard thermal loops are just corrections
$\sim T^2/f_\pi^2$, and except for exceptional circumstances,
do not dominate the terms at tree level.

To illustrate such an exception, consider 
the elastic scattering between two pions, of different
isospin, in the forward direction.
Sitting in the center of mass frame of the two incident pions, 
with euclidean conventions the momenta 
for the process $1^a 2^b\rightarrow3^c4^d$ are 
$P_1=(i p,\vec{p})= - P_3$ and $P_2= (i p,-\vec{p})=-P_4$;
the superscripts denote the isospin indices, $a=c \neq b=d$.  
The original lagrangian
${\cal L}_2$ generates four pion scattering amplitudes
$\sim P^2/f_\pi^2$, with hard thermal loops 
from $\delta {\cal L}_2$ down by $T^2/f_\pi^2$.
For this special choice of momenta and isospin, however,
the term from ${\cal L}_2$, 
$= \delta^{a c} \delta^{b d} (P_1 + P_3)^2/f_\pi^2$, vanishes,
and the leading term is from $\delta{\cal L}_2$, 
$= \frac{2}{9}\delta^{a c} \delta^{b d} (T^2/f_\pi^2) (p^2/f_\pi^2)$.  
In this way, hard thermal loops generate novel scattering
amplitudes between Goldstone bosons at nonzero temperature;
perhaps this might be observable in spin waves.

We can also use these techniques to compute the leading
corrections at low temperature to anomalous
processes.  Purely hadronic anomalous processes are
only allowed when $N \geq 3$, and are
described by the Wess-Zumino-Witten term\cite{wzw}.
This can be written as a lagrangian in five dimensions,
\begin{equation}
{\cal L}_{5} =
\frac{i N_c}{240 \pi^2} \;
\epsilon^{\alpha \beta \delta \gamma \kappa} \;
tr(R_\alpha R_\beta R_\delta R_\gamma R_\kappa) \; ,
\label{e10}
\end{equation}
where $N_c$ is the number of colors, $N_c=3$ in $QCD$,
and $\epsilon^{\alpha \beta \delta \gamma \kappa}$ is
the completely antisymmetric tensor.  Up to an overall
integer $= N_c$, the coefficient of ${\cal L}_5$ is determined by topology
in five dimensions \cite{wzw}.  At the quark level, ${\cal L}_5$
arises from diagrams related to the quark anomaly.
In terms of the chiral fields, we call a term anomalous
if it is odd under the transformation
of $g \rightarrow g^\dagger$; such terms have abnormal
intrinsic parity \cite{wzw}.
The original ${\cal L}_2$ is not anomalous, ${\cal L}_5$ is.

After the redefinition of $g$
in (\ref{e1a}), and using the complete equations of
motion including ${\cal L}_5$,
$\int d^5 x {\cal L}_5 \rightarrow \int d^5 x {\cal L}_5 +
\int d^4 x {\cal L}_5(\pi)$, where 
${\cal L}_5(\pi)$ is a lagrangian density
in four dimensions,
\begin{equation}
{\cal L}_5(\pi) = \frac{i N_c}{24 \pi^2 f_\pi^2} 
\epsilon^{\alpha \beta \delta \gamma} \;
tr \left( \partial_\alpha \pi 
\{ \left[\pi, R_\beta \right], 
R_\delta R_\gamma \} \right) ,
\label{e11}
\end{equation}
and $\{\cdot , \cdot \}$ denotes the anticommutator.
We then compute the anomalous diagram between four $R_\mu$'s at one
loop order, 
using ${\cal L}_5(\pi)$ at one vertex and ${\cal L}_2(\pi)$ at the other.

Once the commutator and anticommutator 
in the vertex of (\ref{e11}) 
are contracted with the $f^{a b c}$ from the
commutator of (\ref{ea4}), the matrix structure simplifies
significantly.  It is natural to introduce another
kind of vector potential,
$A_\mu$, and a field strength, $G_{\mu \alpha}$,
\begin{equation}
A_\mu = \epsilon^{\mu \beta \delta \gamma} \;
R_\beta R_\delta R_\gamma \;\;\; , \;\;\;
G_{\mu \alpha} = \partial_\mu A_\alpha
- \partial_\alpha A_\mu \; .
\label{e12}
\end{equation}
Unlike $R_\mu$, $A_\mu$ is neither a pure
gauge transformation, nor does it transform as a gauge
field under $SU(N)$ gauge transformations.  We
introduce the abelian part of the field
strength for $A_\mu$, $G_{\mu \alpha}$, 
because of (\ref{e7}).  

The integral which arises is again $\Pi^{\mu \nu}(P)$.
Keeping only the terms $\sim T^2$, 
as $tr(A_\mu R_\mu) = 0$,
the term $\sim \delta^{\mu \nu}$ in (\ref{e6}) doesn't contribute
to the anomalous amplitude between four $R_\mu$'s,
while that $\sim \delta \Pi^{\mu \nu}$ does.
Following (\ref{e9}), we then generalize this amplitude 
between four $R_\mu$'s to
write the dominant corrections to the anomalous effective
lagrangian as $\delta{\cal L}_5$,
\begin{equation}
\delta {\cal L}_5 = 
- \frac{i N_c T^2}{144 \pi^2 f_\pi^2} 
\int \frac{d \Omega_{ \hat{k} } }{4 \pi} \; 
tr \left(\! 
G_{\mu \alpha}
\frac{ \hat{K}^\alpha \hat{K}^\beta}{- (D \cdot \hat{K})^2}
F_{\mu \beta}
\!\! \right) .
\label{e13}
\end{equation}
Even though $G_{\mu \alpha}$
does not transform covariantly under $SU(N)$ gauge transformations,
since the leading anomalous terms are given by a single insertion
of $G_{\mu \alpha}$, we can still use the $SU(N)$ gauge invariance
to sum all one loop diagrams with 
one insertion of ${\cal L}_5(\pi)$, and any number of 
cubic vertices from ${\cal L}_2(\pi)$, by replacing
$\partial \cdot \hat{K}$ with $D \cdot \hat{K}$, 
and $\partial_\mu R_\beta - \partial_\beta R_\mu$ with $2 F_{\mu \beta}$,
in (\ref{e13}).

Expanding in terms
of pion fluctuations, as before 
the field strength $F_{\mu \nu}$ is quadratic in the $\pi$'s,
while $G_{\mu \nu} \sim \partial_{ [ \mu , } 
\epsilon_{ \nu ] \alpha \beta \delta}
\partial^\alpha \pi \partial^\beta \pi \partial^\delta \pi/f_\pi^3$
is cubic.  Thus both ${\cal L}_5$ and
$\delta{\cal L}_5$ contribute to a five-point interaction,
$KK \rightarrow \pi\pi\pi$ \cite{wzw}.
Counting powers of momenta,
$\delta {\cal L}_5 \sim P^4(1/P^2)P^2 \sim P^4$
is of the same order as $\int dx_5 {\cal L}_5 \sim P^4$.  

Although $\delta {\cal L}_2$
alters the coefficient of $tr(R_\mu^2)$ in ${\cal L}_2 + \delta {\cal L}_2$,
$\delta{\cal L}_5$ doesn't affect that of ${\cal L}_5$.
Since the normalization of ${\cal L}_5$ is 
fixed by topology in five dimensions, this is natural;
the fluctuations, like ${\cal L}_5(\pi)$, are always four dimensional,
and so only produce corrections in four dimensions,
such as $\delta {\cal L}_5$.
The same thing happens at zero temperature\cite{ren}.

This approach can also be used to compute the temperature
corrections to the chiral lagrangian including the coupling
to external gauge fields\cite{pt2}.  This exercise is
interesting in its own right; for example, we find that
the amplitude for $\pi^0 \rightarrow 2 \gamma$ decreases
to $\sim T^2/f_\pi^2$.  This is in accord with the
behavior near the chiral phase transition,
where this amplitude vanishes \cite{rp1}.  

To conclude, cool pions give us a broader perspective 
on hard thermal loops.
For example, at next to leading order 
about zero temperature, $\sim T^4/f_\pi^4$, besides
corrections to the nonlocal terms in $\delta{\cal L}_2$ and
$\delta{\cal L}_5$, perhaps there are also new,
nonlocal terms.  This suggests that the known hard
thermal loops, as discussed herein, 
are themselves the first terms in an infinite series
of nonlocal effective lagrangians at nonzero temperature.  

This work is supported by a DOE grant at 
Brookhaven National Laboratory, DE-AC02-76CH00016.

\end{narrowtext}

\begin{references}
\bibitem{foot}
If instead of (\ref{e1a}) we define $g \rightarrow u (exp(i \pi/f_\pi)) u$
and expand about a background field $u$,
then the covariant derivative which appears in the ensuing
${\cal L}_2(\pi)$ is
$D_\mu = \partial_\mu + \frac{1}{2}[V_\mu,\cdot]$, where
$V_\mu = u^\dagger \partial_\mu u + u \partial_\mu u^\dagger$.
Viewed as an $SU(N)$ gauge field, $V_\mu$ is
not a pure gauge transformation, and so the field strength 
$F_{\mu \nu} = 
\partial_\mu V_\nu - \partial_\nu V_\mu + \frac{1}{2} [V_\mu, V_\nu]$
is nontrivial.  With this parametrization, the effective
lagrangian ${\cal L}_2 + \delta {\cal L}_2$ 
of (\ref{e9}) is given by taking $R_\mu = (u^\dagger)^2 \partial_\mu
u^2$ in $tr(R_\mu^2)$, and this
$D_\mu$ and $F_{\mu \nu}$ in the nonlocal term.
While this form of the effective lagrangian appears very
different, on the mass shell it should give the same
scattering amplitudes as (\ref{e9}).
\end{references}
\end{document}